**Coaxial Nanowire Resonant Tunneling Diodes from non-polar AlN/GaN on Silicon**

S. D. Carnevale[1], C. Marginean[2], P. J. Phillips[1], T. F. Kent[1], A. T. M. G. Sarwar[3], M. J. Mills[1], R. C. Myers[1,2,3,a]

[1]Department of Materials Science and Engineering, The Ohio State University, Columbus, Ohio, 43210, USA

[2]Department of Physics, The Ohio State University, Columbus, OH, 43210, USA

[3]Department of Electrical and Computer Engineering, The Ohio State University, Columbus, Ohio, 43210, USA

**Abstract**

**Resonant tunneling diodes are formed using AlN/GaN core-shell nanowire heterostructures grown by plasma assisted molecular beam epitaxy on n-Si(111) substrates. By using a coaxial geometry these devices take advantage of non-polar (m-plane) nanowire sidewalls. Device modeling predicts non-polar orientation should enhance resonant tunneling compared to a polar structure and that AlN double barriers will lead to higher peak-to-valley current ratios compared to AlGaN barriers. Electrical measurements of ensembles of nanowires show negative differential resistance appearing only at cryogenic temperature. Individual nanowire measurements show negative differential resistance at room temperature with peak current density of $5\times10^5$ A/cm$^2$.**

Resonant tunneling diodes (RTDs) are important unipolar quantum electronic devices that display a strong bias induced change in current when electrons resonantly tunnel through double barriers within the device's conduction band. Negative differential resistance (NDR) results when an increase in voltage raises the injected electron energy above the resonant tunneling condition, leading to a decrease in current due to reduced probability of tunneling. RTDs are attractive for use in THz devices[1-3] and multi-valued logic circuits[4] because of their high-speed capabilities and non-linear IV characteristics. RTDs operating at fundamental frequencies above 1 THz at room temperature are now possible[5, 6]. While the majority of work to date has focused on III-V RTDs such as GaAs/AlGaAs[7, 8] and InAs/AlSb[9], AlN and GaN exhibit a large conduction band offset (2.1eV)[10], high electron mobilities, and good thermal stability that could enhance RTD performance.

---

[a] Author to whom correspondence should be addressed. Electronic mail: myers.1079@osu.edu.

Several groups have previously reported NDR in double-barrier AlGaN/GaN RTDs grown along polar directions[11-14]. The NDR in these structures is commonly hysteretic; it disappears after repeated scans and it is not present when scanning backwards (i.e. from V > 0 to V = 0), which has been explained as due to tunneling and charging of trap states[12, 13]. Low Al composition AlGaN/GaN RTDs formed along non-polar directions have also been grown[15] and show more reliable and reproducible NDR than their polar equivalents. Recently Sakr et al. proposed that the hysteric NDR reported in AlN/GaN heterostructures was not due to resonant tunneling, but from an effect of charging and discharging of trap states[16]. They argue that because this NDR has been predominately observed at room temperature and disappeared at low temperatures it cannot be due to tunneling.

Here we report on GaN/AlN RTDs formed using core-shell catalyst-free nanowire (NW) heterostructures grown on silicon wafers. The coaxial geometry provides an m-plane (nonpolar) quantum well for ideal flatband tunneling. At cryogenic temperatures clear NDR features are observed, strongly indicating that the transport mechanism is tunneling. Individual NW measurements carried out at room temperature also display NDR, with a peak current density of $5 \times 10^5$ A/cm$^2$.

The AlN/GaN double barrier structure is grown in a core-shell NW geometry (see Fig. 1) without catalysts by plasma assisted molecular beam epitaxy (PAMBE)[17]. A two-step growth method was used to grow Si doped (nominally $1\times10^{19}$ cm$^{-3}$) n-GaN NWs at low densities[17] (~1-10 μm$^{-2}$) on n-Si(111) substrates. NWs are first nucleated at 765°C until they reach the desired areal density, then the substrate temperature is increased to 780°C. At the higher growth temperature no new NWs nucleate and the NWs that have already nucleated continue to grow vertically, thus maintaining a low density. These NWs act as the n-GaN cores for the heterostructure. Once the NW cores have reached a desired length the substrate temperature is reduced to 500°C to promote coaxial formation of GaN and the double barrier structure (Figure 1b) consisting of 1.5nm of i-AlN/2.5nm i-GaN/1.5 nm i-AlN/15 nm n-GaN ($1\times10^{19}$ cm$^{-3}$ Si). The band diagram (Fig. 1c) for this structure is modeled using a one-dimensional, self-consistent Schrodinger-Poisson solver[18]. It is assumed that the Fermi level is pinned 0.55eV below the conduction band at the sidewall surface based on previous measurements of surface states on GaN NW sidewalls[19].

The nonpolar (m-plane) structures provide several important advantages stemming from the flat band configuration. A one-dimensional non-equilibrium Green's function based model is developed, using the commercial software ATLAS, to simulate RTDs with polar surfaces, non-polar surfaces, and different barrier heights (AlGaN %Al) (Fig. 2). Tunneling occurs at smaller electron energies for a non-polar structure when compared to a polar structure (Fig. 2a). Simulations also predict that increasing the barrier height from AlGaN to AlN should lead to a higher peak-to-valley current ratio (PVCR) (Fig. 2b). In a planar device, AlN barriers will reduce material quality due to the incorporation of strain related defects, but AlN barriers can be included in NWs because they can accommodate strain without generating extended defects[20].

Furthermore, III-nitride NWs grown by PAMBE are generally attractive because their material quality is exceptionally good[21] and they grow on Si substrates, allowing for integration into current Si-based technologies. The effectiveness of NWs for this application was previously shown in a vertically aligned AlN/GaN NW RTD[22]. However, in the case of coaxial devices, the current density (normalized by the device planar area) will scale with NW height since the device active region is a cylindrical NW shell. Thus, in principle, NW coaxial RTDs could exhibit higher total current (per planar area) than a planar RTD.

Electron microscopy is used to study the structural characteristics of the NWs. Scanning electron microscopy (SEM) images of the sample in cross-section and plan view are provided in Fig. 3a and b, respectively. All SEM images are obtained on an FEI Sirion scanning electron microscope operating at an accelerating voltage between 10 to 20 kV using the "in lens" secondary electron detector. The thick layer of material grown between NWs on the Si substrate forms during deposition at low substrate temperatures. The large bulb of faceted material at the top of the NWs is due to decreased Ga adatom mobility at low substrate temperatures[17].

Individual NW heterostructures are analyzed using high resolution scanning transmission microscopy (STEM). Images of single NWs obtained using an FEI Titan[3] 80-300 probe-corrected monochromated scanning/transmission electron microscope operated at 300 kV are displayed in Fig. 3c and d. High-angle annular dark field (HAADF) conditions provide Z contrast (GaN is light, AlN is dark). The image in Fig. 3c is taken from a device grown with a thinner outermost shell than the structure shown in 3d. The thinner shell clearly reduces the amount of the faceted material at the NW top. However, a thicker shell may be necessary to prevent depletion of the outermost shell due to surface states. Atomic resolution images (Fig. 3e) show that the layer thicknesses of the active region agree with the target device design, though the outermost shell is somewhat thinner than intended. As expected, each NW is also found by STEM to be free of extended defects. Additionally, the bulb at the top of the NW is mainly composed of vertical deposition of the same layer structure on the sidewalls, but each vertical layer is approximately ten times thicker than each coaxial layer. Thus tunneling through the 12 nm thick AlN vertical barrier instead of the 1.5 nm thin AlN coaxial barrier should be insignificant. However, the interface between the vertical and coaxial material offers potential current leakage pathways. This could be avoided in future devices by planarizing the NWs in order to electrically contact only the coaxial NW regions.

Electrical measurements of ensembles of NWs are performed at both room temperature and 11 K (Fig. 4a). An insulating layer is formed in the area between the NWs by applying a spin on glass (Hydrogen silsesquioxane or HSQ) that is then cured at 350°C for 30 minutes. A 20 nm Ti/50 nm Au contact that covers both the NW tops and sidewalls is then deposited. The insulating layer prevents the metal contact from creating a short to the Si substrate, while the thicker vertical AlN barriers ensure that the dominant current pathway is coaxial. A second contact is formed directly to the Si substrate by first removing the NWs from the surface and then annealing a small piece of In onto the bare Si surface. A positive potential is applied to the NW contact while the substrate is grounded. Current-voltage scans are carried out piecewise in small intervals (1-2V) to better see NDR characteristics and allow for immediate repetition of measurements to look for hysteresis. At room temperature no NDR is observed (Fig. 4c). However, when the sample is cooled to 11K in vacuum, NDR is apparent in both the initial and second scan (Fig. 4d), but disappears by the third. PVCRs of 1.6 and 1.9 are found for this device. The fact that the NDR is not present in ensemble measurements at room temperature but is present at cryogenic temperatures is consistent with a tunneling mechanism leading to NDR.

Single NW electrical measurements are performed at room temperature using a FEI Helios Nanolab 600 Dual Beam Focused Ion Beam/Scanning Electron Microscope equipped with Kleindiek nanomanipulators capable of making electrical contact to individual nanowires. A nanomanipulator probe is placed at the top of a NW to bias its shell and a second probe is placed directly onto the Si substrate (Fig. 4b). Because the number of NWs contacted is known, the current density can be normalized using the area of the NW's cross-section. A representative room temperature current density vs. voltage (J-V) characteristic for a single NW RTD displaying clear NDR is shown in Fig. 4e. Figure 4e also shows a measurement of two NWs simultaneously contacted by the same nanoprobe. Given the difficulty in maintaining contact to a single NW in this measurement scheme it was challenging to repeatedly measure any identical NW without damaging it. In the small number of cases in which this was possible (not shown here) the NDR disappears after repeated measurements and is not present in backwards scans (i.e. positive bias to zero bias). We also note that single NW nanoprobe measurements can be distorted by the applied stress of the nanoprobe. The measurement shown for a single NW has a PVCR of 2.4. A peak resonant tunneling current density of $5\times10^5 A/cm^2$ is found, roughly five times higher than any previously reported current density in an AlN/GaN RTD[14]. This current density is comparable to what has previously been achieved in III-V RTDs[23, 24], though it is smaller than the recently reported current densities greater than $10^6 A/cm^2$.[25]

The experimental data in Fig. 4e are modeled using a simple circuit consisting of three components in series: the ideal RTD as modeled in Fig. 2, a Schottky diode between the top n-GaN region and nanoprobe tip, and a generalized series resistance. The RTD barrier and quantum well are assumed to be 1.25-nm and 2-nm thick, respectively, which is within 1-2 monolayers of the thicknesses measured in the STEM image shown in Fig. 3d. The current density is scaled by the ratio of NW sidewall area to NW top area (~18) to account for the effect of device geometry. The Schottky diode is modeled according to the electron affinity of GaN and the work function of the tungsten probe (4.5eV). The series resistance is adjusted to achieve the best fit possible, corresponding to $4\mu\Omega\text{-cm}^2$. The data and simulation are in remarkably close agreement given that only one fit parameter is varied.

To sum, we fabricated a coaxial (m-plane) nanowire AlN/GaN RTD grown on Si(111) substrates by PAMBE. STEM shows the NWs to be free of extended defects. Ensemble I-V measurements show NDR that only appears at cryogenic temperatures, but that disappears after repeated scans. Individual NW measurements taken at room temperature also show NDR, with a peak current density of $5\times10^5 \text{A/cm}^2$. These results show that resonant tunneling is feasible in nonpolar III-Nitride NWs and reveals a pathway for Nitride-based RTDs integrated on silicon.

**Acknowledgements**

We thank Dr. Patrick Roblin for helpful discussion regarding this work. This research was supported by the Office of Naval Research (N00014-09-1-1153) and by the National Science Foundation CAREER award (DMR-1055164). S.D.C. acknowledges support from the National Science Foundation Graduate Research Fellowship Program (2011101708).

**Figure Captions**

FIG 1. (Color online) (a) Schematic AlN/GaN resonant tunneling diode design (b) Layer compositions within the active region of the device. (c) Band diagram of the device including three quasi-bound states within the GaN quantum well in the conduction band.

FIG 2. (Color online) Results of device modeling. (a) Simulated transmittance as a function of electron energy for non-polar, Ga-polar and N-polar AlN/GaN resonant tunneling diodes (b) Simulated peak-to-valley current ratio as a function of barrier composition.

FIG 3. (Color online) (a) Cross section and (b) plan view scanning electron microscopy images of the as grown nanowire sample. Z-contrast scanning transmission electron microscopy (STEM) images of nanowire resonant tunneling diodes (c) with a thin outer shell and (d) a thick outer shell. (e) Z-contrast atomic resolution STEM image of the device active region.

FIG 4. (Color online) Contact design for measuring the current-voltage characteristics of (a) an ensemble of nanowires and (b) using a nanoprobe to contact a small number of wires (arrows show the position of probes). (c) Room temperature I-V using the scheme shown in (a) (Inset shows I-V in log scale). (d) I-V of an ensemble of nanowires at cryogenic temperatures with room temperature measurements included as reference. (e) Experimentally measured J-V of a single nanowire and two simultaneously contacted nanowires showing NDR. A simulated J-V of a single nanowire (plotted as grey line) is included to compare with experimental results.

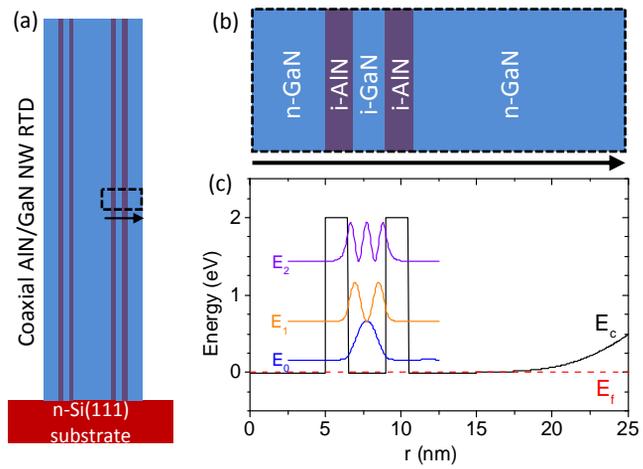

Figure 1, Carnevale et al.

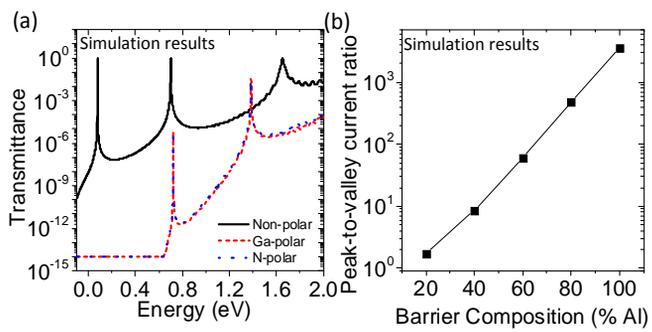

Figure 2, Carnevale et al.

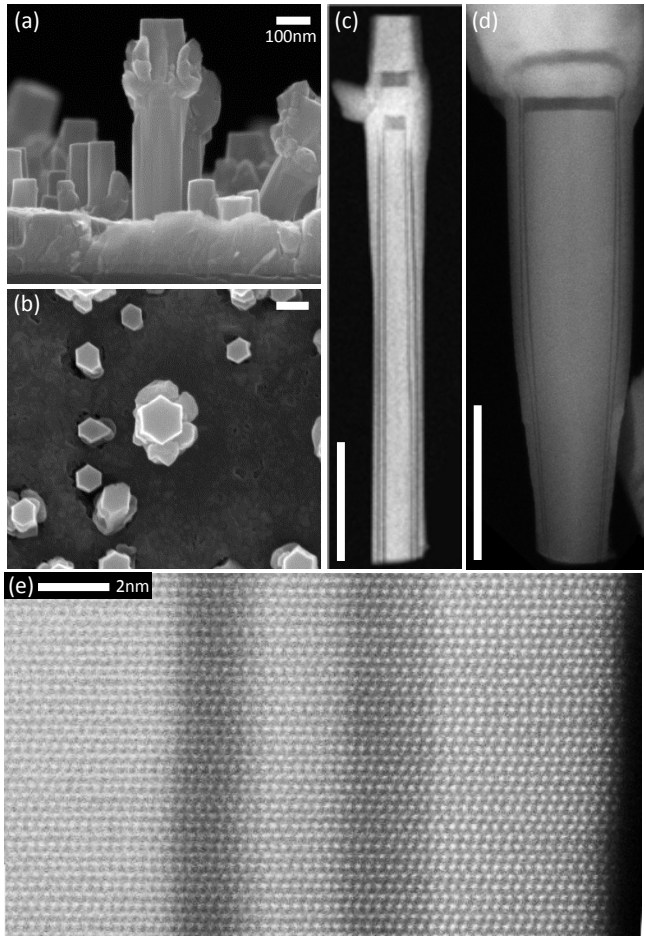

Figure 3, Carnevale et al.

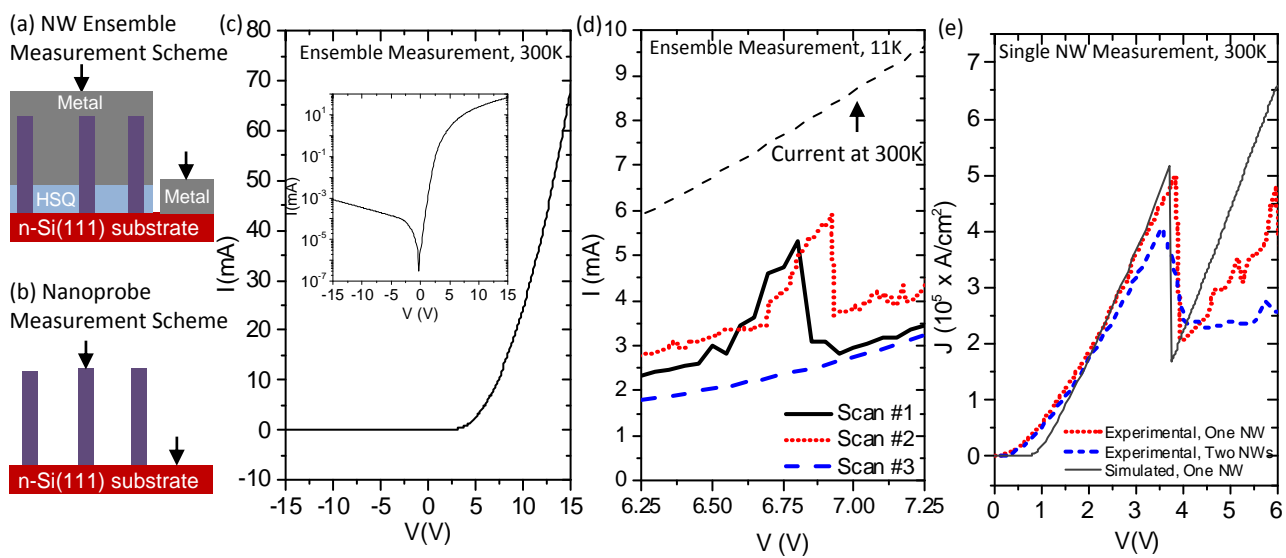

Figure 4, Carnevale et al.